\documentclass{article}
\usepackage{spconf,amsmath,graphicx}
\usepackage{makecell}
\usepackage[table,xcdraw]{xcolor}
\usepackage{multirow}
\usepackage{amssymb}
\usepackage{breqn}

\usepackage{caption}
\usepackage{titlesec}
\titlespacing*{\section}{0pt}{0.2\baselineskip}{0.2\baselineskip}
\titlespacing*{\subsection}{0pt}{0.1\baselineskip}{0.2\baselineskip}
\usepackage{enumitem}
\setlist{nosep}  

\title{Spiking-LEAF: A Learnable Auditory front-end for \\ Spiking Neural Networks}
\name{Zeyang Song$^1$, Jibin Wu$^{2,*}$, Malu Zhang$^3$, Mike Zheng Shou$^1$, Haizhou Li$^{1,4}$ \thanks{$^{*}$Corresponding author: jibin.wu@polyu.edu.hk}\thanks{This work was supported in part by IAF, A*STAR, SOITEC, NXP and National University of Singapore under FD-fAbrICS: Joint Lab for FD-SOI Always-on Intelligent \& Connected Systems (Award I2001E0053), the Agency for Science, Technology and Research (A*STAR) under its AME Programmatic Funding Scheme (Project No. A18A2b0046), and the Research Grants Council of the Hong Kong SAR (Grant No. PolyU25216423), The Hong Kong Polytechnic University (P0043563 and P0046094), The National Natural Science Foundation of China (Grant No. 62306259).}}
\address{$^1$National University of Singapore, Singapore\\
$^2$The Hong Kong Polytechnic University, Hong Kong SAR, China\\
$^3$University of Electronic Science and Technology of China, China \\
$^4$Shenzhen Research Institute of Big Data, School of Data Science, \\ The Chinese University of Hong Kong, Shenzhen, China}
\begin{document}
%
\maketitle
\begin{abstract}
Brain-inspired spiking neural networks (SNNs) have demonstrated great potential for temporal signal processing.  However, their performance in speech processing remains limited due to the lack of an effective auditory front-end. To address this limitation, we introduce Spiking-LEAF, a learnable auditory front-end meticulously designed for SNN-based speech processing. Spiking-LEAF combines a learnable filter bank with a novel two-compartment spiking neuron model called IHC-LIF. The IHC-LIF neurons draw inspiration from the structure of inner hair cells (IHC) and they leverage segregated dendritic and somatic compartments to effectively capture multi-scale temporal dynamics of speech signals. Additionally, the IHC-LIF neurons incorporate the lateral feedback mechanism along with spike regularization loss to enhance spike encoding efficiency. On keyword spotting and speaker identification tasks, the proposed Spiking-LEAF outperforms both SOTA spiking auditory front-ends and conventional real-valued acoustic features in terms of classification accuracy, noise robustness, and encoding efficiency.  
\end{abstract}
%
\section{Introduction}
\label{sec:intro}

Recently, the brain-inspired spiking neural networks (SNNs) have demonstrated superior performance in sequential modeling \cite{yin2021accurate, ijcai2023p0344}. However, their performance in speech processing tasks still lags behind that of state-of-the-art (SOTA) non-spiking artificial neural networks (ANNs) \cite{bittar2022surrogate,wu2020deep,wu2018spiking,wu2018biologically,pan2019neural,zhang2020efficient,wu2021progressive,Xingyi22ECCV,Xinyin23NeurIPS}. This is primarily due to the lack of an effective auditory front-end that can synergistically perform acoustic feature extraction and neural encoding with high efficacy and efficiency. 


The existing SNN-based auditory front-ends first extract acoustic features from raw audio signals, followed by encoding these real-valued acoustic features into spike patterns that can be processed by the SNN. For feature extraction, many works directly adopt the frequently used acoustic features based on the Mel-scaled filter-bank \cite{bittar2022surrogate,wu2020deep,wu2018spiking} or the GammaTone filter-bank~\cite{pan2020efficient}. Despite the simplicity of this approach, these handcrafted filter-bank are found to be suboptimal in many tasks when compared to learnable filter-bank \cite{sainath2015learning, hoshen2015speech,ravanelli2018speaker,zeghidour2021leaf}.
In another vein of research, recent works have also looked into the neurophysiological process happening in the peripheral auditory system and developed more complex biophysical models to enhance the effectiveness of feature extraction \cite{cramer2020heidelberg,legenstein2008learning}. However, these methods not only require fine-tuning a large number of hyperparameters but are also computationally expensive for resource-constrained neuromorphic platforms. 

For neural encoding, several methods have been proposed that follow the neurophysiological processes within the cochlea \cite{cramer2020heidelberg, legenstein2008learning}. For instance, Cramer et al. proposed a biologically inspired cochlear model with the model parameters directly taken from biological studies \cite{cramer2020heidelberg}. Additionally, other methods propose to encode the temporal variations of the speech signals that are critical for speech recognition. The Send on Delta (SOD)~\cite{miskowicz2006send} and threshold coding methods~\cite{pan2020efficient,zhang2019mpd, schrauwen2003bsa}, for instance, encode the positive and negative variations of signal amplitude into spike trains. However, these neural encoding methods lack many essential characteristics as seen in the human's peripheral auditory system that are known to be important for speech processing, such as feedback adaptation \cite{bear2020neuroscience}.  


To address these limitations, we introduce a Spiking LEarnable Audio front-end model, called Spiking-LEAF. The Spiking-LEAF leverages a learnable auditory filter-bank to extract discriminative acoustic features. Furthermore, inspired by the structure and dynamics of the inner hair cells (IHCs) within the cochlea, we further proposed a two-compartment neuron model for neural encoding, namely IHC-LIF neuron. Its two neuronal compartments work synergistically to capture the multi-scale temporal dynamics of speech signals. Additionally, the lateral inhibition mechanism along with spike regularization loss is incorporated to enhance the encoding efficiency. The main contributions of this paper can be summarized as follows:
\begin{itemize}
\item We propose a learnable auditory front-end for SNNs, enabling the joint optimization of feature extraction and neural encoding processes to achieve optimal performance in the given task.
\item We propose a two-compartment spiking neuron model for neural encoding, called IHC-LIF, which can effectively extract multi-scale temporal information with high efficiency and noise robustness.
\item Our proposed Spiking-LEAF shows high classification accuracy, noise robustness, and encoding efficiency on both keyword spotting and speaker identification tasks.

\end{itemize}

\begin{figure}
    \centering
    \includegraphics[width=0.8\linewidth]{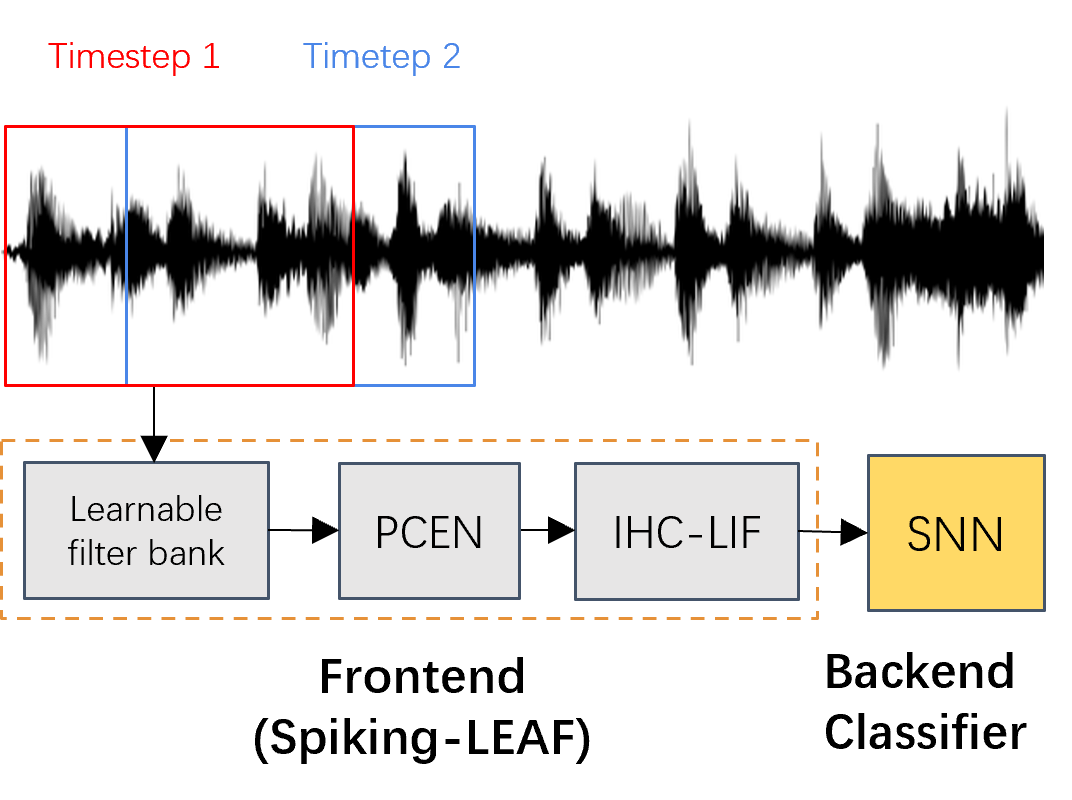}
    \caption{The overall architecture of the proposed SNN-based speech processing framework.}
    \label{frontend diagram}
\end{figure}

\section{Methods}

As shown in Fig. \ref{frontend diagram}, similar to other existing auditory front-ends, the proposed Spiking-LEAF model consists of two parts responsible for feature extraction and neural encoding, respectively. For feature extraction, we apply the Gabor 1d-convolution filter bank along with the Per-Channel Energy Normalization (PCEN) to perform frequency analysis. Subsequently, the extracted acoustic feature is processed by the IHC-LIF neurons for neural encoding. Given that both the feature extraction and neural encoding parts are parameterized, they can be optimized jointly with the backend SNN classifier. 

\subsection{Parameterized acoustic feature extraction}

In Spiking-LEAF, the feature extraction is performed with a 1d-convolution Gabor filter bank along with the PCEN that is tailored for dynamic range compression~\cite{wang2017trainable}. The Gabor 1d-convolution filters have been widely used in speech processing~\cite{zeghidour2018learning, zeghidour2021leaf}, and its formulation can be expressed as per:
\begin{equation}
\begin{split}
    \phi_ {n}  (t)=  e^ {i2\pi \eta _ {n}t}& \frac{1}{\sqrt {2\pi }\sigma _ {n}}  e^ {-\frac {t^ {2}}{2\sigma _ {n}^ {2}}}
\end{split}
\end{equation}
where $\eta_n$ and $\sigma_n$ denote learnable parameters that characterize the center frequency and bandwidth of filter n, respectively. In particular, for input audio with a sampling rate of 16 kHz, there are a total of 40 convolution filters, with a window length of 25ms ranging over $t=-L/2, ..., L/2$ ($L=401$ samples), have been employed in Spiking-LEAF. 
These 1d-convolution filters are applied directly to the audio waveform $x$ to get the time-frequency representation $F$. 

Following the neurophysiological process in the peripheral auditory system, the PCEN \cite{zeghidour2021leaf,wang2017trainable} has been applied subsequently to further compress the dynamic range of the obtained acoustic features:
\begin{align}
&PCEN(F(t,n))=\left(\frac {F(t,n)}{(\varepsilon +M(t,n))^ {\alpha _ {n}}+\delta _ {n}}\right)^{r_n} - \delta _ {n}^ {r_ {n}} \\
&M(t,n)=(1-s)M(t-1,n)+sF(t,n)
\end{align}

In Eqs. 2 and 3, $F(t,n)$ represents the time-frequency representation for channel $n$ at time step $t$. $r_n$ and $\alpha_{n}$ are coefficients that control the compression rate. The term $M(t,n)$ is the moving average of the time-frequency feature with a smoothing rate of $s$. Meanwhile, $\varepsilon$ and $\delta_{n}$ stands for a positive offset introduced specifically to prevent the occurrence of imaginary numbers in PCEN. 

\subsection{Two-compartment spiking neuron model}
\begin{figure}
    \centering
    \includegraphics[width=\linewidth]{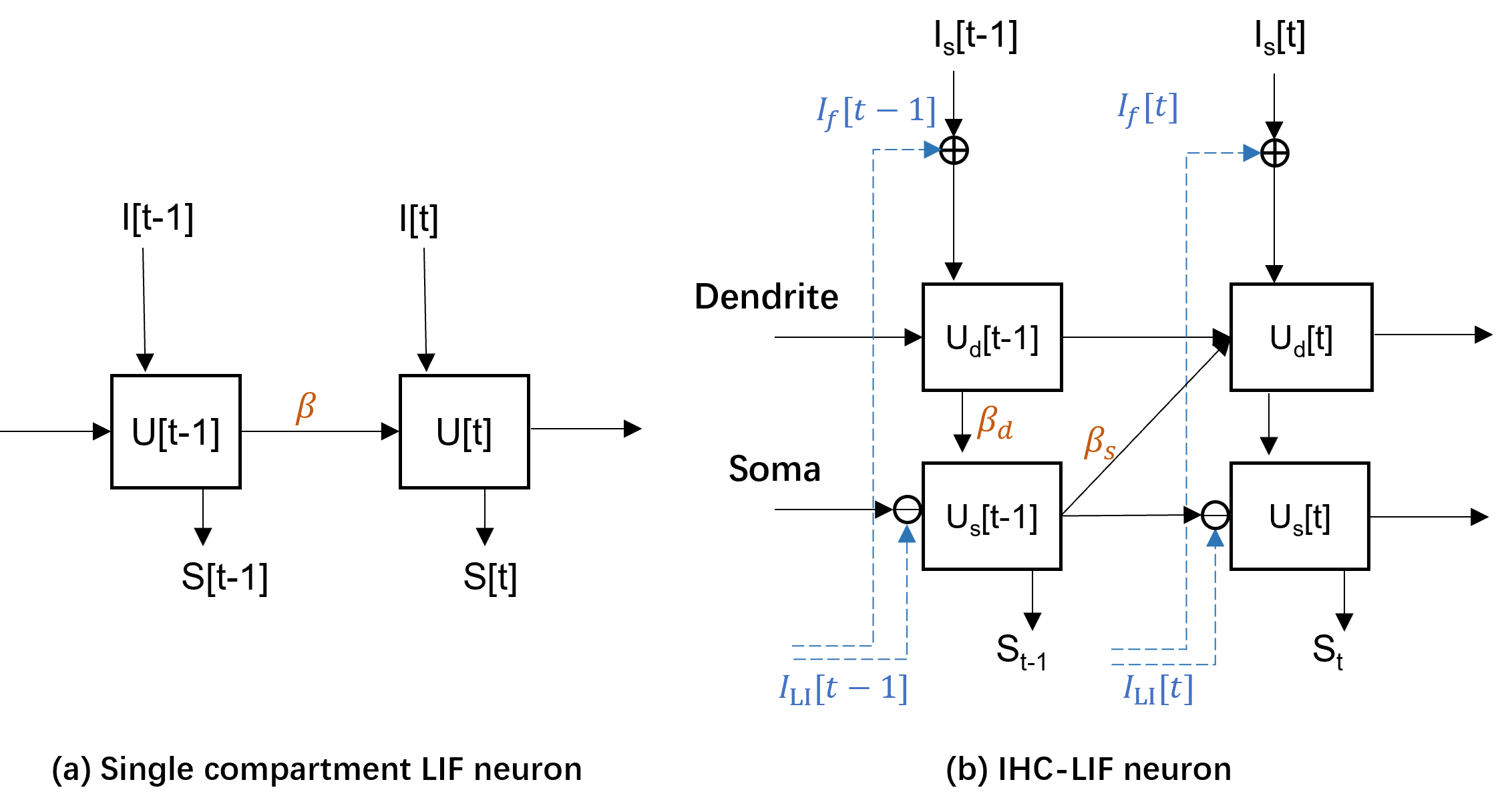}
    \caption{Computational graphs of LIF and IHC-LIF neurons. }
    \label{IHC-LIF}
\end{figure}

The Leaky Integrate-and-Fire (LIF) neuron model \cite{gerstner2002spiking}, with a single neuronal compartment, has been widely used in brain simulation and neuromorphic computing \cite{bittar2022surrogate,wu2020deep,wu2018spiking,pan2019neural,zhang2020efficient}. The internal operations of a LIF neuron, as illustrated in Fig. \ref{IHC-LIF} (a), can be expressed by the following discrete-time formulation:
\begin{align}
    &I[t] = \Sigma_i w_iS[t-1] + b\\
    &U[t] = \beta * U[t-1] +  I[t] - V_{th}S[t-1]\\
    \label{LIF_thres} &S[t] = \mathbb{H}(U[t] - V_{th})
\end{align}
where $S[t-1]$ represents the input spike at time step $t$. $I[t]$ and $U[t]$ denote the transduced synaptic current and membrane potential, respectively. $\beta$ is the membrane decaying constant that governs the information decaying rate within the LIF neuron. As the Heaviside step function indicated in Eq. \ref{LIF_thres}, once the membrane potential exceeds the firing threshold $V_{th}$, an output spike will be emitted.

Despite its ubiquity and simplicity, the LIF model possesses inherent limitations when it comes to long-term information storage. These limitations arise from two main factors: the exponential leakage of its membrane potential and the resetting mechanism. These factors significantly affect the model's efficacy in sequential modeling. Motivated by the intricate structure of biological neurons, recent work has developed a two-compartment spiking neuron model, called TC-LIF, to address the limitations of the LIF neuron~\cite{zhang2023long}. The neuronal dynamics of TC-LIF neurons are given as follows:
\begin{align}
    I[t] &= \Sigma_i w_iS[t-1] + b\\
    \begin{split}
    U_d[t] &= U_d[t-1] + \beta_d * U_s[t-1] + I[t] \\ 
    &\quad - \gamma * S[t-1]
    \end{split}\\
    U_s[t] &= U_s[t-1] + \beta_s * U_d[t-1] - V_{th}S[t-1]  \\
    S[t] &= \mathbb{H}(U[t] - V_{th})
\end{align}
where $U_d[t]$ and $U_s[t]$ represent the membrane potential of the dendritic and somatic compartments. The $\beta_d$ and $\beta_s$ are two learnable parameters that govern the interaction between dendritic and somatic compartments. Facilitated by the synergistic interaction between these two neuronal compartments, TC-LIF can retain both short-term and long-term information which is crucial for effective speech processing~\cite{zhang2023long}. 

\subsection{IHC-LIF neurons with lateral feedback}
Neuroscience studies reveal that lateral feedback connections are pervasive in the peripheral auditory system, and they play an essential role in adjusting frequency sensitivity of auditory neurons~\cite{hudspeth2014integrating}. Inspired by this finding, as depicted in Figure \ref{IHC-LIF} (b), we further incorporate lateral feedback components into the dendritic compartment and somatic compartment of the TC-LIF neuron, represented by $I_{f}[t]$ and $I_{LI}[t]$ respectively. Specifically, each output spike will modulate the neighboring frequency bands with learnable weight matrices $ZeroDiag(W_f)$ and $ZeroDiag(W_{LI})$, whose diagonal entries are all zeros. 

The lateral inhibition feedback of hair cells within the cochlea is found to detect sounds below the thermal noise level and in the presence of noise or masking sounds~\cite{guinan2018olivocochlear, sasmal2018competition}. Motivated by this finding, we further constrain the weight matrix $W_{LI}\geq0$ to enforce lateral  inhibitory feedback at the somatic compartment, which is responsible for spike generation. This will suppress the activity of neighboring neurons after the spike generation, amplifying the signal of the most activated neuron while suppressing other neurons. This results in a sparse yet informative spike representation of input signals. The neuronal dynamics of the resulting IHC-LIF model can be described as follows: 
\begin{align}
    I_{s}[t] &= \Sigma_i w_iS[t-1] + b\\
    I_{f}[t] &= ZeroDiag(W_f) * S[t-1]\\
    I_{LI}[t] &= ZeroDiag(W_{LI}) * S[t-1]\\
    \begin{split}
    U_d[t] &= U_d[t-1] +\beta_d * U_s[t-1] + I_s[t] \\
    &\quad  - \gamma * S[t-1] + I_f[t]
    \end{split}\\
    \begin{split}
    U_s[t] &= U_s[t-1] + \beta_s * U_d[t-1] -V_{th} S[t-1]) \\
    &\quad  - I_{LI}[t]
    \end{split}\\
    S[t] &= \mathbb{H}(U_s[t] - V_{th})
\end{align}
To further enhance the encoding efficiency, we incorporate a spike rate regularization term $L_{SR}$ into the loss function $L$. It has been applied alongside the classification loss $L_{cls}:L = L_{cls} + \lambda L_{SR}$ where $L_{SR} = ReLU(R - SR)$. Here, $R$ represents the average spike rate per neuron per timestep and $SR$ denotes the expected spike rate. Any spike rate higher than $SR$ will incur a penalty, and $\lambda$ is the penalty coefficient. 

\begin{table}[]
\resizebox{\linewidth}{!}{\begin{tabular}{llccc}
\hline
Tasks& Front-end& \begin{tabular}[c]{@{}c@{}}Classifier \\ Structure\end{tabular} & \begin{tabular}[c]{@{}c@{}}Classifier \\ Type\end{tabular} & \begin{tabular}[c]{@{}c@{}}Test \\ Accuracy (\%)\end{tabular} \\ \hline
                    & Fbank \cite{bittar2022surrogate} & 512-512 & Feedforward & 83.03\\
                    & Fbank+LIF& 512-512& Feedforward&85.24\\
                    &Heidelberg\cite{cramer2020heidelberg}& 512-512 & Feedforward & 68.14 \\
                    &\textbf{Spiking-LEAF}&512-512 &Feedforward &\textbf{92.24} \\ 
                    &Speech2spike \cite{stewart2023speech2spikes} & 256-256-256 & Feedforward & 88.5 \\
                    &\textbf{Spiking-LEAF}&256-256-256 &Feedforward &\textbf{90.47} \\ \cline{2-5}
                    & Fbank \cite{bittar2022surrogate}& 512-512 & Recurrent & 93.58 \\
                    & Fbank+LIF& 512-512& Recurrent&92.04\\
\multirow{-9}{*}{KWS}& \textbf{Spiking-LEAF} &512-512 &Recurrent &\textbf{93.95} \\ \hline
                             &Fbank& 512-512& Feedforward&29.42\\
                             &Fbank+LIF& 512-512& Feedforward&27.23\\
                             &\textbf{Spiking-LEAF}&512-512& Feedforward&\textbf{30.17}\\\cline{2-5}
                             &Fbank& 512-512& Recurrent&31.76\\
                             &Fbank+LIF& 512-512& Recurrent&29.74\\ 
\multirow{-6}{*}{SI} & \textbf{Spiking-LEAF}&512-512& Recurrent&\textbf{32.45} \\ \hline
\end{tabular}}
\caption{Comparison of different auditory front-ends on KWS and SI tasks. The bold color denotes the best model for each network configuration.}
\label{Tasks}
\end{table}

\section{Experimental Results}
In this section, we evaluate our model on keyword spotting (KWS) and speaker identification task. For keyword spotting, we use Google Speech Command Dataset V2 \cite{speechcommandsv2}, which contains 105,829 one-second utterances of 35 commands.  For speaker identification (SI), we use the Voxceleb1 dataset~\cite{Nagrani17} with 153,516 utterances from 1,251 speakers, resulting in a classification task with 1,251 classes. We focus our evaluations on the auditory front-end by keeping model architecture and hyperparameters of the backend SNN classifier fixed.

\subsection{Superior feature representation}

Table \ref{Tasks} compares our proposed Spiking-LEAF model with other existing auditory front-ends on both KWS and SI tasks. Our results reveal that the Spiking-LEAF consistently outperforms the SOTA spike encoding methods as well as the fbank features~\cite{bittar2022surrogate}, demonstrating a superior feature representation power. In the following section, we validate the effectiveness of key components of Spiking-LEAF: learnable acoustic feature extraction, two-compartment LIF~(TC-LIF) neuron model, lateral feedback $I_f$, lateral inhibition $I_{LI}$, and firing rate regulation loss $L_{SR}$.

\subsection{Ablation studies}
\textbf{Learnable filter bank and two-compartment neuron.} As illustrated in row 1 and row 2 of Table \ref{spike rate}, the proposed learnable filter bank achieves substantial enhancement in feature representation when compared to the widely adopted Fbank feature. Notably, further improvements in classification accuracy are observed (see row 3) when replacing LIF neurons with TC-LIF neurons that offer richer neuronal dynamics. However, it is important to acknowledge that this improvement comes at the expense of an elevated firing rate, which has a detrimental effect on the encoding efficiency.

\begin{table}[]
\centering
\resizebox{\linewidth}{!}{%
\begin{tabular}{ccccccc}
\hline
Acoustic features& Neuron type     & $I_f$& $I_{LI}$ & $L_{SR}$ & Firing rate & Accuracy \\ \hline
Fbank& LIF & - & -  & -     & 17.94\%    & 85.24\%  \\
Learnable& LIF  & -   & -  & -     & 18.25\%    & 90.73\%  \\
Learnable& TC-LIF  & -   & -  & -     & 34.21\%    & 91.89\%  \\
Learnable& TC-LIF  & \checkmark   & -  & -     & 40.35\%    & 92.24\%  \\
Learnable& TC-LIF & \checkmark   & \checkmark  & -     & 34.54\%    & \textbf{92.43\%}  \\
Learnable& TC-LIF  & \checkmark   & -  & \checkmark     & 15.03\%    & 90.82\%  \\
Learnable& TC-LIF  & \checkmark  & \checkmark  & \checkmark     & \textbf{11.96\%}    & 92.04\%  \\ 
\hline

\end{tabular}%
}
\vspace{2pt}
\caption{Ablation studies of various components of the proposed Spiking-LEAF model on the KWS task. 
}
\label{spike rate}
\end{table}

\begin{figure}
    \centering
    \includegraphics[width=\linewidth]{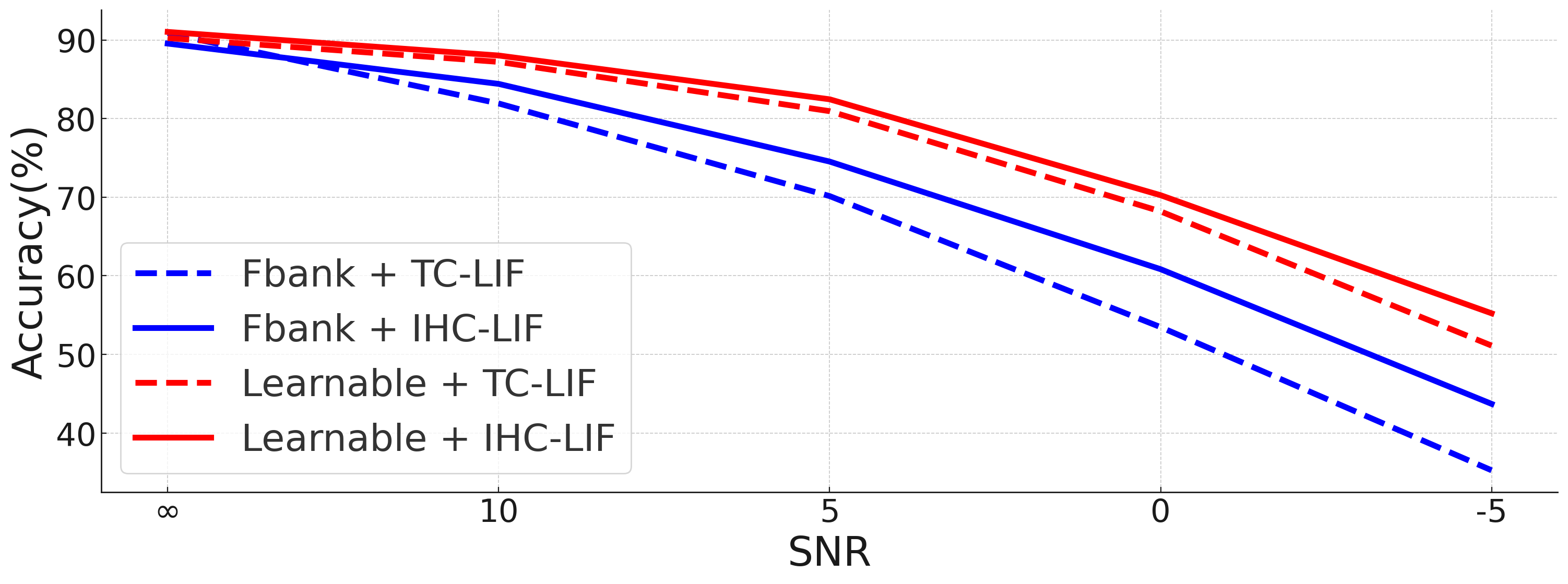}
    \caption{Test accuracy on the KWS task with varying SNRs.
    }
    \label{SNR comp}

\end{figure}

\begin{figure}
    \centering
    \includegraphics[width=0.83\linewidth]{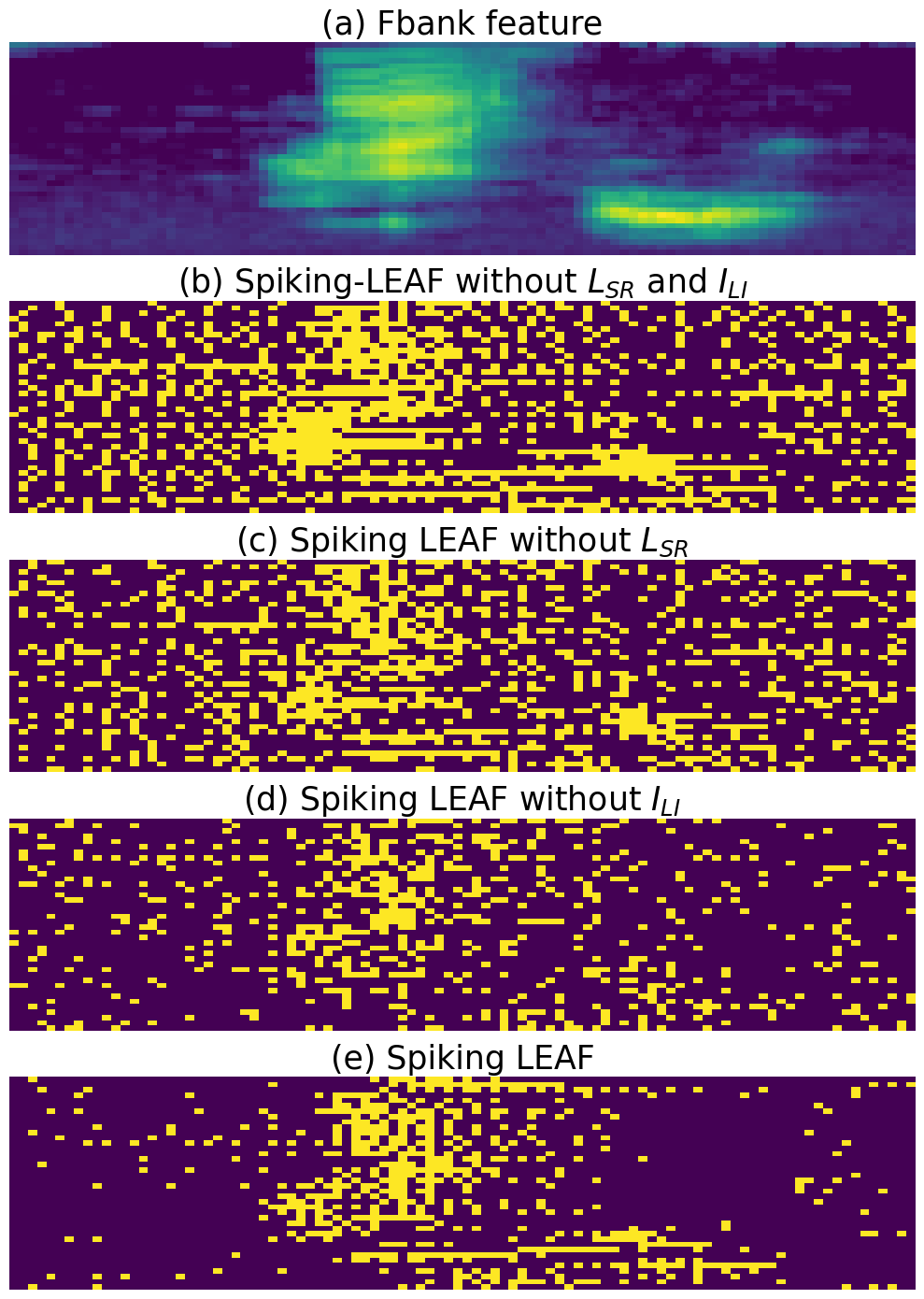}
    \vspace{2pt}
    \caption{This figure illustrates the  Fbank feature and spike representation generated by Spiking-LEAF without and with lateral inhibition and spike rate regularization loss.}
    \label{Spike regularization}
\end{figure}

\textbf{Lateral feedback.} Row 4 and row 5 of Table \ref{spike rate} highlight the potential of lateral feedback mechanisms in enhancing classification accuracy, which can be explained by the enhanced frequency sensitivity facilitated by the lateral feedback. Furthermore, the incorporation of lateral feedback is also anticipated to enhance the neuron's robustness in noisy environments. To substantiate this claim, our model is trained on clean samples and subsequently tested on noisy test samples contaminated with noise from the NOISEX-92 \cite{varga1993assessment} and CHiME-3 \cite{barker2015third} datasets. Fig. \ref{SNR comp} illustrates the results of this evaluation, demonstrating that both the learnable filter bank and lateral feedback mechanisms contribute to enhanced noise robustness. This observation aligns with prior studies that have elucidated the role of the PCEN in fostering noise robustness \cite{zeghidour2021leaf}. Simultaneously, Fig. \ref{Spike regularization} showcases how the lateral feedback aids in filtering out unwanted spikes.

\textbf{Lateral inhibition and spike rate regularization loss.} As seen in Fig. \ref{Spike regularization} (b), when the spike regularization loss and lateral inhibition are not applied, the output spike representation involves a substantial amount of noise during non-speech periods. Introducing lateral inhibition or spike regularization loss alone can not suppress the noise that appeared during such periods (Figs. (b) and (c)). Particularly, introducing the spike regularization loss alone results in a uniform reduction in the output spikes (Fig. \ref{Spike regularization} (d)). However, this comes along with a notable reduction in accuracy as highlighted in Table \ref{spike rate} row 6. Notably, the combination of lateral inhibition and spike rate regularization (Fig. \ref{Spike regularization} (e)) can effectively suppress the unwanted spike during non-speech periods, yielding a sparse and yet informative spike representation.

\section{Conclusion}
\label{sec:conclusion}
In this paper, we presented a fully learnable audio front-end for SNN-based speech processing, dubbed Spiking-LEAF. The Spiking-LEAF integrated a learnable filter bank with a novel IHC-LIF neuron model to achieve effective feature extraction and neural encoding. Our experimental evaluation on KWS and SI tasks demonstrated enhanced feature representation power, noise robustness, and encoding efficiency over SOTA auditory front-ends. It, therefore, opens up a myriad of opportunities for ultra-low-power speech processing at the edge with neuromorphic solutions. 

\vfill
\pagebreak

\bibliographystyle{IEEEbib}
\footnotesize
\newpage
\bibliography{refs}

\end{document}